\documentclass[apj]{emulateapj}
\usepackage{amsmath}

\shorttitle{SAMI: Galaxy Interactions and Kinematic Anomalies in Abell 119}
\shortauthors{Oh et al.}

\def\deg{^\circ}
\def\sur{\rm mag\,\, arcsec^{-2}}

\begin{document}

\title{The SAMI Galaxy Survey: Galaxy interactions and Kinematic Anomalies in Abell 119}

\author{Sree Oh$^{1}$ }
\author{Sukyoung K. Yi$^{1}$ }
\author {Luca Cortese$^{2}$}
\author {Jesse van de Sande$^{3}$}
\author {Smriti Mahajan$^{4}$}
\author {Hyunjin Jeong$^{5}$}
\author {Yun-Kyeong Sheen$^{5}$}
\author {James T. Allen$^{3,6}$}
\author {Kenji Bekki$^{2}$}
\author {Joss Bland-Hawthorn$^{3}$}
\author {Jessica V. Bloom$^{3,6}$}
\author {Sarah Brough$^{7}$}
\author {Julia J. Bryant$^{3,6,7}$}
\author {Matthew Colless$^{6,8}$}
\author {Scott M. Croom$^{3,6}$}
\author {L. M. R. Fogarty$^{3,6}$}
\author {Michael Goodwin$^{7}$}
\author {Andy Green$^{7}$}
\author {Iraklis S. Konstantopoulos$^{7,9}$}
\author {Jon Lawrence$^{7}$}
\author { \'A. R. L\'opez-S\'anchez\altaffilmark{7,10}}
\author {Nuria P. F. Lorente$^{7}$}
\author {Anne M. Medling$^{8,11,12}$}
\author {Matt S. Owers$^{7,10}$}
\author {Samuel Richards$^{3,6,7}$}
\author {Nicholas Scott$^{3,6}$}
\author {Rob Sharp$^{8}$}
\author {Sarah M. Sweet$^{8}$}

\affil{$^1$ Department of Astronomy \& Yonsei University Observatory, Yonsei University, Seoul 120-749, Korea; \\ sree@galaxy.yonsei.ac.kr; yi@yonsei.ac.kr\\
$^2$ ICRAR M468, University of Western Australia, 35 Stirling Highway, Crawley WA 6009, Australia
$^3$ Sydney Institute for Astronomy (SIfA), School of Physics, The University of Sydney, NSW 2006, Australia\\
$^4$ Indian Institute of Science Education and Research Mohali, IISERM, Punjab, India\\
$^5$ Korea Astronomy and Space Science Institute, Daejeon 305-348, Korea\\
$^6$ ARC Centre of Excellence for All-sky Astrophysics (CAASTRO)\\
$^7$ Australian Astronomical Observatory PO Box 915, North Ryde NSW 1670, Australia\\
$^8$ Research School of Astronomy and Astrophysics Australian National University, Canberra ACT 2611, Australia\\
$^{9}$ Envizi Group Suite 213, National Innovation Centre, Australian Technology Park, 4 Cornwallis Street, Eveleigh NSW 2015, Australia\\
$^{10}$ Department of Physics and Astronomy, Macquarie University, NSW 2109, Australia\\
$^{11}$ Cahill Center for Astronomy and Astrophysics California Institute of Technology, MS 249-17, Pasadena, CA 91125, USA\\
$^{12}$ Hubble Fellow}

\submitted{Accepted 7 September 2016}

\begin{abstract}
Galaxy mergers are important events that can determine the fate of a galaxy by changing its morphology, star-formation activity and mass growth. Merger 
systems have commonly been identified from their disturbed morphologies, and we now can employ Integral Field Spectroscopy to detect and analyze the impact of mergers on stellar kinematics as well. We visually classified galaxy morphology using deep images ($\mu_{\rm r} = 28\,\sur$) taken by the Blanco 4-m telescope at the Cerro Tololo Inter-American Observatory. In this paper we investigate 63 bright ($M_{\rm r}<-19.3$) spectroscopically-selected galaxies in Abell 119; of which 53 are early type and 20 galaxies show a disturbed morphology by visual inspection. A misalignment between the major axes in the photometric image and the kinematic map is conspicuous in morphologically-disturbed galaxies. Our sample is dominated by early-type galaxies, yet it shows a surprisingly tight Tully-Fisher relation except for the morphologically-disturbed galaxies which show large deviations. Three out of the eight slow rotators in our sample are morphology disturbed. The visually-selected morphologically-disturbed galaxies are generally more asymmetric, visually as well as kinematically. Our findings suggest that galaxy interactions, including mergers and perhaps fly-bys, play an important role in determining the orientation and magnitude of galaxy's angular momentum. 
\end{abstract}

\keywords{catalogs -- galaxies: interactions -- galaxies: kinematics and dynamics --
galaxies: evolution -- galaxies: clusters: individual Abell 119}

\section{Introduction}
\label{sec:intro}

Galaxies experience the most dramatic changes during galaxy mergers. Mergers can sometimes change the galaxy morphology (Toomre \& Toomre 1972; Barnes \& Hernquist 1992), generate starbursts (Cox et al. 2008; Kaviraj 2010a;  L\'opez-S\'anchez 2010), stimulate nuclear activity (Canalizo et al. 2007; Bennert et al. 2008; Ellison et al. 2011; Cotini et al. 2013; Satyapal et al. 2014), and contribute to stellar mass growth (Lee \& Yi 2013). Therefore, merger history of galaxies is important to understand present-day galaxies. Since mergers cause distortions in the morphology of the system, visual inspections of images are often used to trace recent mergers. Deep images are therefore essential for detecting evidence of mergers and interactions. Deep images reveal that galaxies have assembled substantial fraction of their mass via mergers even in recent epochs (Van Dokkum 2005; Tal et al. 2009; Kaviraj et al. 2010b; Sheen et al. 2012; Yi et al. 2013; Duc et al. 2015).

Kinematics plays a key role in determining the morphology of galaxies, and therefore, visually observable disturbances are also linked to the underlying kinematics (Rampazzo et al. 2005; Kronberger et al. 2007; Yang et al. 2008). Integral field spectroscopy (IFS) allows us to access 2D spatially-resolved velocity fields, so the analyses that are used to detect features in an image can be applied to IFS data as well. As we visually define peculiar morphologies using images, kinematically distinct features (e.g., kinematically distinct cores, counter-rotating cores, and double peaks in velocity dispersion) can be detected  using 2D kinematic maps of galaxies (e.g., Krajnovi$\acute{\rm c}$ et al. 2011).  

Several techniques quantifying peculiarities in gas and stellar kinematics have been applied to large datasets from various surveys (Shapiro et al. 2008; Krajnovi$\acute{\rm c}$ et al. 2011). A recent study by Bloom et al. (2016, submitted) shows that morphologically disturbed galaxies correlate with asymmetry in gas kinematics better than ``stellar'' (i.e., photometric) parameters, such as Gini (Abraham et al. 2003), M20 (Lotz et al. 2004), or CAS (Abraham et al. 1996; Conselice 2003). The distribution of gas is more easily disturbed than that of stars, making gas an efficient tracer of kinematic disturbances (Stil \& Israel 2002, Cannon et al. 2004, Koribalski \&  L\'opez-S\'anchez 2009,  L\'opez-S\'anchez et al. 2012). Gas however cannot be used for gas-deficient early-type galaxies. We therefore use stellar kinematics for our cluster galaxy sample that is predominantly early type and gas deficient.

Kinematic studies from IFS surveys have changed our view of galaxy morphology. The SAURON (de Zeeuw et al. 2002) and ATLAS$^{\rm 3D}$ surveys (Cappellari et al. 2011) reported that galaxies can be separated into two categories, fast and slow rotators, according to their spin parameter ($\lambda_{\rm R}$). In their classification, the majority of early-type galaxies show fast and regular rotation, which implies that they have measurable levels of 
 ordered motion (e.g., Emsellem et al. 2007; Cappellari et al. 2007; Emsellem et al. 2011; Krajnovi$\acute{\rm c}$ et al. 2013). This however is 
 counterintuitive to the traditional idea of a ``pressure-supported'' early-type and ``ordered-motion supported'' late-type description of galaxies, thereby 
 questioning the origin and evolution of angular momentum in different types of galaxies. 

The investigation of mergers on angular momentum changes can provide clues to the origin of slow rotators. 
Numerical simulations suggest that galaxies can lose their angular momentum via mergers and become slow rotators (Bois et al. 2011; Khochfar et al. 2011; Naab et al. 2014, and references therein).
To empirically assess the role of mergers on the evolution of angular momentum, it is necessary to know the intrinsic distribution of angular momentum of stable disks, before a galaxy undergoes an interaction. The Tully-Fisher  relation (T-F, Tully \& Fisher 1977) serves this purpose. The tight correlation between stellar luminosity and HI line width has been well established by many studies on disk galaxies in the local Universe. Furthermore, the T-F relation is found even among early-type galaxies using HI and CO line observations (Williams, Bureau \& Cappellari 2010; Davis et al. 2011; Heijer et al. 2011). Recently, Cortese et al. (2014) found that early-type galaxies exhibit T-F relation similar to spirals, although with markedly larger scatter. These findings imply that the T-F relation is a general trend in ``regular'' rotating systems regardless of galaxy morphology; and if dynamical interactions change the angular momentum of galaxies, they introduce scatter in the T-F relation (Kassin et al. 2007; Covington et al. 2010; De Rossi et al. 2012).

Several studies from numerical simulations suggest that the direction of angular momentum can be easily changed via gas accretion from cosmic webs or galaxy mergers (e.g. Kimm et al. 2011; Sharma, Steinmetz, \& Bland-Hawthorn 2012; Dubois et al. 2014). The misalignment of orientation between images and stellar kinematics was first introduced in Franx, Illingworth \& de Zeeuw (1991), and after that several IFS surveys also detected the kinematic misalignment. Krajnovi$\acute{\rm c}$ et al. (2011) reported that 10\% of early-type galaxies, some of which show prominent merger features, show misaligned kinematic and morphological orientations. Using data from the CALIFA survey (S{\'a}nchez et al. 2012), Barrera-Ballesteros et al. (2015) compared kinematic misalignment of interacting and undisturbed galaxies, and found large misalignments among galaxies that seem to be on-going mergers. On the other hand, the misaligned galaxies in their sample do not always show merger signatures, and misalignment was more often found among galaxies that were rounder or had low angular momentum, making it difficult to conclude on the effect of mergers on kinematic anomalies. It is therefore necessary to use deeper imaging data that reveal the merger status of galaxies clearly to answer this question. 

Here we investigate the impact of mergers on three aspects of kinematics: angular momentum, orientation, and asymmetry. In order to do this we used deep optical images obtained with the Blanco 4-m telescope at the Cerro Tololo Inter-American Observatory (CTIO, Sheen et al. 2012), and stellar kinematics from the Sydney-AAO Multi-object Integral field (SAMI, Croom et al. 2012) galaxy survey (Bryant et al. 2015). Specifically, we quantitatively compare kinematic properties between morphologically-disturbed and undisturbed galaxies in Abell 119 and discuss the merger origin of their kinematic anomalies.
Throughout the paper, we adopt a standard $\Lambda$CDM cosmology with $\Omega_{\rm m} = 0.3$, $\Omega_\Lambda= 0.7$, and $H_0 = 70$ km/s/Mpc.

\section[]{Data and Samples}

\label{sec:data}
 
  \subsection{CTIO Deep imaging}
  
 We obtained deep optical images of Abell 119 in $u$, $g$, and $r$ bands using the MOSAIC II CCD mounted on the Blanco telescope. These observations were carefully planned with the goal of detecting low surface brightness features related to galaxy interactions. For Abell 119, seven dithered exposures were taken and combined to make an $r$-band stacked image with a total integration time of 5,040 seconds. The large field of view (FOV) of MOSAIC II CCD ($36' \times 36'$) covers $0.7\,R_{200}$ of Abell 119 ($\sim$1.53 Mpc) at $z =$ 0.044. The surface brightness profiles of galaxies go down to 30 magnitude per square arcsecond in the $r$ band, which allows us to detect extremely faint features ($\mu_{\rm r} = 28\ \sur$, 1$\sigma$ above the sky level). Magnitudes in this study are corrected for foreground extinction using the reddening maps from Schlegel et al. (1998). Details of data reduction and treatment of photometric data are described in Sheen et al. (2012).

 \subsection{The SAMI Galaxy Survey}
 The SAMI instrument (Croom et al. 2012) is mounted on the 3.9-m Anglo-Australian Telescope (AAT) and fed into the AAOmega spectrograph (Sharp et al. 2006). It has 13 `hexabundles' of 15$''$ diameter (Bland-Hawthorn et al. 2011; Bryant et al. 2014); each is comprised of 61 1.6$''$-diameter optical fibers. The hexabundles patrol an FOV of 1-degree diameter. The beam is split into blue (580 V grating, 3700-5700 $\AA$) and red (1000R grating, 6300-7400 $\AA$) arms; the spectral resolution of the blue and red gratings are R = 1730 and R = 4500, respectively. The main SAMI survey will target 3,700 nearby galaxies at 0.04 $< z <$ 0.095 from the Galaxy And Mass Assembly (GAMA; Driver et al. 2011) to obtain an unbiased sample of stellar mass (Bryant et al. 2015) when completed. In addition, SAMI contains eight rich clusters at  $z <$ 0.1 (EDCC0442, Abell 85, Abell 119, Abell 168, Abell 2399, Abell 3880, APMCC0917, and Abell 4038) to cover the high-density environments missed by the main survey. 
 
 The raw data were reduced using the ${\rm{2DFDR}}$ package (Croom et al. 2004) which carries out bias subtraction, flat fielding, wavelength calibration, and sky subtraction. The pairs of data cubes (blue and red) for each galaxy were built after applying telluric correction and flux calibration. See Sharp et al. (2015) and Allen et al. (2015) for a full description of the data reduction. The stellar line-of-sight velocity distribution and velocity dispersion for all galaxies  were derived using the pPXF code (Cappellari \& Emsellem 2004) following van de Sande et al. (in prep.), and two-dimensional stellar kinematic maps consisting of $0.5''$ resampled spatial pixels (spaxels) were built from the cubes. In this study, we used stellar velocity fields without binning with a signal-to-noise (S/N) cutoff of 5 for the continuum in each spaxel.

 \subsection{The sample}   
  Cluster membership of Abell 119 was determined by the redshift information from the Sloan Digital Sky Survey (SDSS; York et al. 2000; Abazajian et al. 2009), 2-degree Field Galaxy Redshift Survey (2dFGRS; Colless et al. 2001), and Hydra on CTIO (Sheen et al. 2012). Galaxies with $M_{*} >10^{9.5} M_{\odot}$ and $R<R_{200}$ were selected as primary targets for SAMI observations for this cluster. The SAMI cluster observations are described in detail in Owers et al. (in prep.). Of the 254 primary targets in Abell 119, 108 galaxies were observed by SAMI by the end of 2015. Our final sample contains 63 galaxies brighter than $r = 17$ ($M_{\rm r}= -19.3$) after cross-matching the 108 galaxies observed by SAMI with the deep images. Our sample contains both early- and late-type galaxies because the parent targets for SAMI were selected regardless of galaxy morphology (or color). Figure 1 shows six sample galaxies. The gray scale background with contours shows the deep image, the colored isophotes are for the specific values of surface brightness as given in the caption, and the colored map at the center shows the SAMI stellar velocity map. 

    \begin{figure*}[p] 
    \centering
    \includegraphics[width=0.85\textwidth]{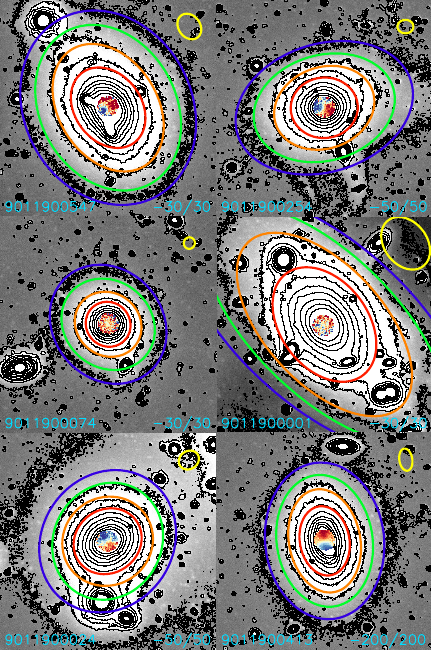}
    \caption[Sample galaxies of Abell 119]
     {This figure shows sample deep images (160$''$ $\times$ 160$''$) from CTIO with the SAMI stellar velocity map with a S/N cutoff of 5/spaxel. The solid lines indicate the isophotal ellipses at $\mu_{\rm r}=$24 (red), 25 (orange), 26 (green), and 27 (blue) $\sur$. The ellipse in the top right corner shows the size of $R_{\rm e}$. The SAMI id of each galaxy is shown in the bottom left. The color range of stellar velocity map is optimized for each galaxy, and the minimum/maximum velocity of each galaxy is indicated in the bottom right.}
    \label{cmd}
    \end{figure*}


\section[]{Analysis}
\subsection{Detecting disturbed features}
 
    \begin{figure}[tb]
    \centering
    \includegraphics[width=0.48\textwidth]{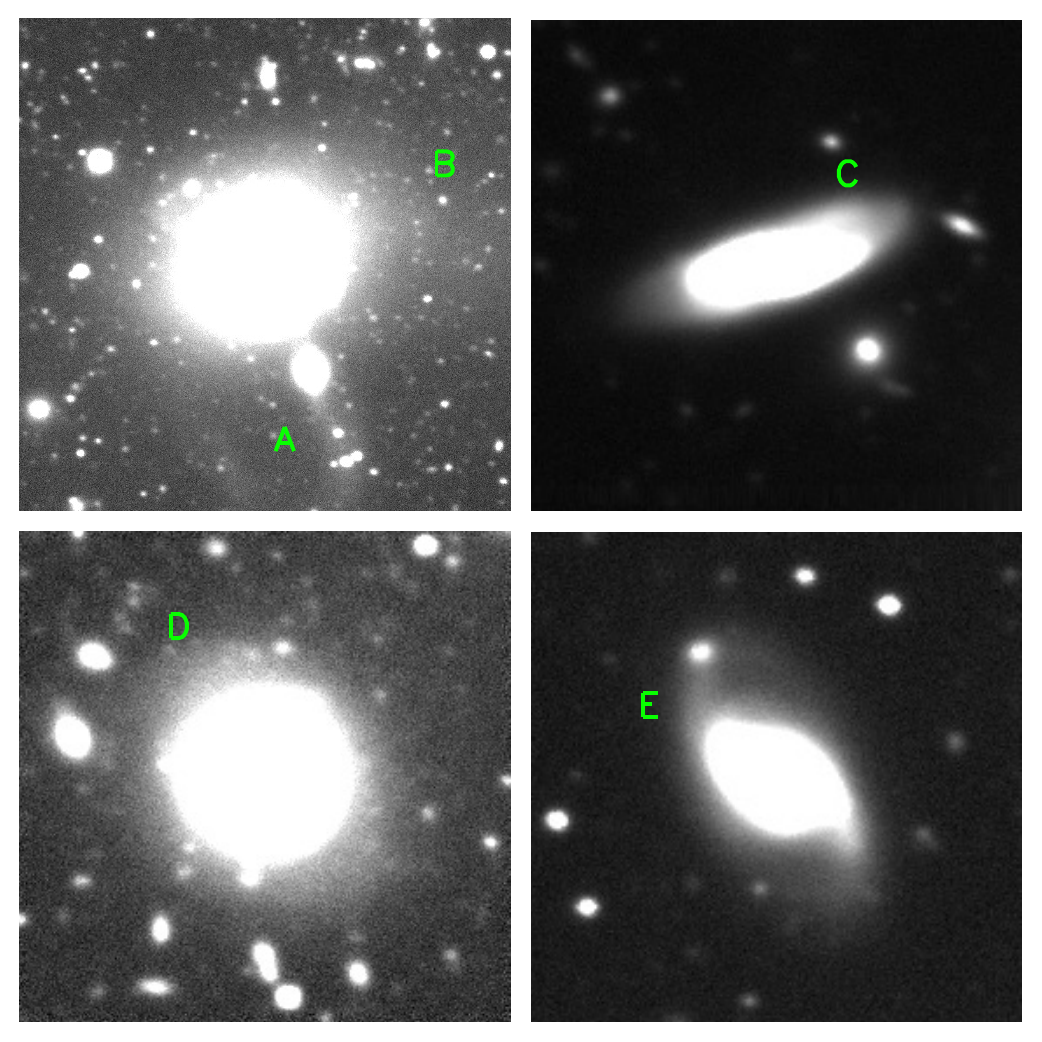}
    \caption[Sample images of morphologically-disturbed galaxies]
     {Sample images of morphologically-disturbed galaxies in Abell 119. Top left: the galaxy shows a tidal tail (A) and asymmetric light distribution (B). Top right: the late-type galaxy is warped and has an unusual strip of light across the galaxy (C). Bottom left: the galaxy displays extended light with irregular shape (D). Bottom right: The galaxy shows a tidal tail (E). Deep images of all sample with a note of our visual classification are provided in Appendix.}
    \label{img}
    \end{figure}

 We visually selected galaxies that showed evidence for interactions or mergers in the $r$-band deep images. Of the 63 galaxies in our sample, 20 galaxies displayed disturbed morphology of varying degree (Figure~\ref{img}). We considered galaxies as morphologically-disturbed when they displayed disturbed features including tidal structure, asymmetric light profile, or unusual dust lanes. If galaxies had close-projected companions, we did not classify them as morphologically-disturbed galaxies unless they showed any of the above-mentioned distortion features. This cluster has been studied for the visual selection of post-merger features in red sequence galaxies by Sheen et al. (2012). In their study using the same images (but for a larger number (133) of member galaxies), they found that approximately 30$\%$ of red sequence galaxies show post-merger features, which is comparable to this study (32$\%$, 20/63). Abell 119 has been imaged by SDSS; therefore, we can directly compare our classifications of the same galaxies at different image qualities. For comparison, we could find only 5 disturbed galaxies using the same visual classification on SDSS images. We present the images of all 63 galaxies in the Appendix, but it is necessary to inspect the images with varying scale to properly determine the level of disturbance. The identification of morphologically-disturbed galaxies is admittedly subjective, and we further discuss this in Section 5.2.
 
We classify the 63 galaxies in our sample into 53 early- (E/S0) and 10 late-type galaxies.
Four of the 20 morphologically-disturbed galaxies are late type. Considering the high early-type fraction and the dominance of the brightest cluster galaxy (BCG), this puts A119 into a cD cluster category (Way, Quintana, \& Infante 1997). Its dominance with early-type galaxies within $0.7\,R_{200}$ is also consistent with the typical morphology-density relation found in clusters (Dressler et al. 1997). Abell 119 is found to be a dynamically young cluster having unrelaxed groups within the Virial radius (Lee et al. 2016). Its dynamical mass has been estimated to be $2.6\,\times\,10^{14}\,M_{\odot}$ (Sheen et al. 2012).

\subsection{Photometry and fitting}
  We measured the effective radius ($R_{\rm e}$), position angle ($PA$), and ellipticity ($\epsilon = 1-a/b$) using the $ellipse$ task in IRAF\footnote{IRAF is distributed by  NOAO which is operated by AURA Inc., under cooperative agreement with NSF} STSDAS. The center of the galaxies was fixed; however, we allowed free estimation of the position angle and ellipticity at a given level of light. All the contaminants were masked out to consider only the light from the target galaxy. We used the curve of growth of light from the radial intensity profile to measure the effective (half-light) radius $R_{\rm e}$. 

\subsection{Kinemetric analysis}
 We measured the spatially-resolved stellar velocity observed with SAMI using the $kinemetry$ procedure (Krajnovi$\acute{\rm c}$ et al. 2006). The $kinemetry$ procedure calculates the expansion of the surface photometry on velocity fields with high-order moments:
\begin{equation}
{K(a,\psi) = A_0(a) + \sum_{n=1}^N(A_n(a)sin(n\psi)+B_n(a)cos(n\psi))}
\end{equation}
where $a$ is the semi-major radius of each kinematic ellipse and $\psi$ is the azimuthal angle of the projected major axis. The semi-major axis of each kinematic ellipse was increased by following $a = i + 1.1^i$, where $i$ = 1, 2, 3 to avoid possible correlations between ellipse elements. The number of ellipse elements was varied between 20 and 100 (but typically over 60). We measured the ellipse parameters as long as spaxels cover 75\% area of the ellipse, and the largest ellipse typically reaches $1-2\,R_{\rm e}$ for our sample. The orientation and flattening of the iso-velocity contours are measured along the kinemetric ellipse.

The amplitude coefficients for each kinemetric ellipse ($k_n$) can be calculated using the harmonic coefficients ($A_n$ and $B_n$):
\begin{equation}
k_n = \sqrt{A_n^2+B_n^2}.
\end{equation}
We used 6 terms for the harmonic analysis to obtain 3 odd terms (n=1, 3, 5).
The coefficient $k_1$ describes the amplitude of the circular velocity, and the higher order terms ($k_3$ and $k_5$) represent complex structure of the velocity field, and deviations from the bulk motion. These coefficients are used to calculate the model rotational velocity in Section 4.1 and the kinematic asymmetry in Section 4.3.
For a detailed description of the significance of the harmonic coefficients for a thin disk case the reader is referred to Wong et al. (2004) and references therein.

\section[]{Results}
\label{sec:results}

\subsection{Rotational velocity and $\lambda_{\rm R}$}
\label{res:vel}
    \begin{figure}[tb]
    \centering
    \includegraphics[width=0.5\textwidth]{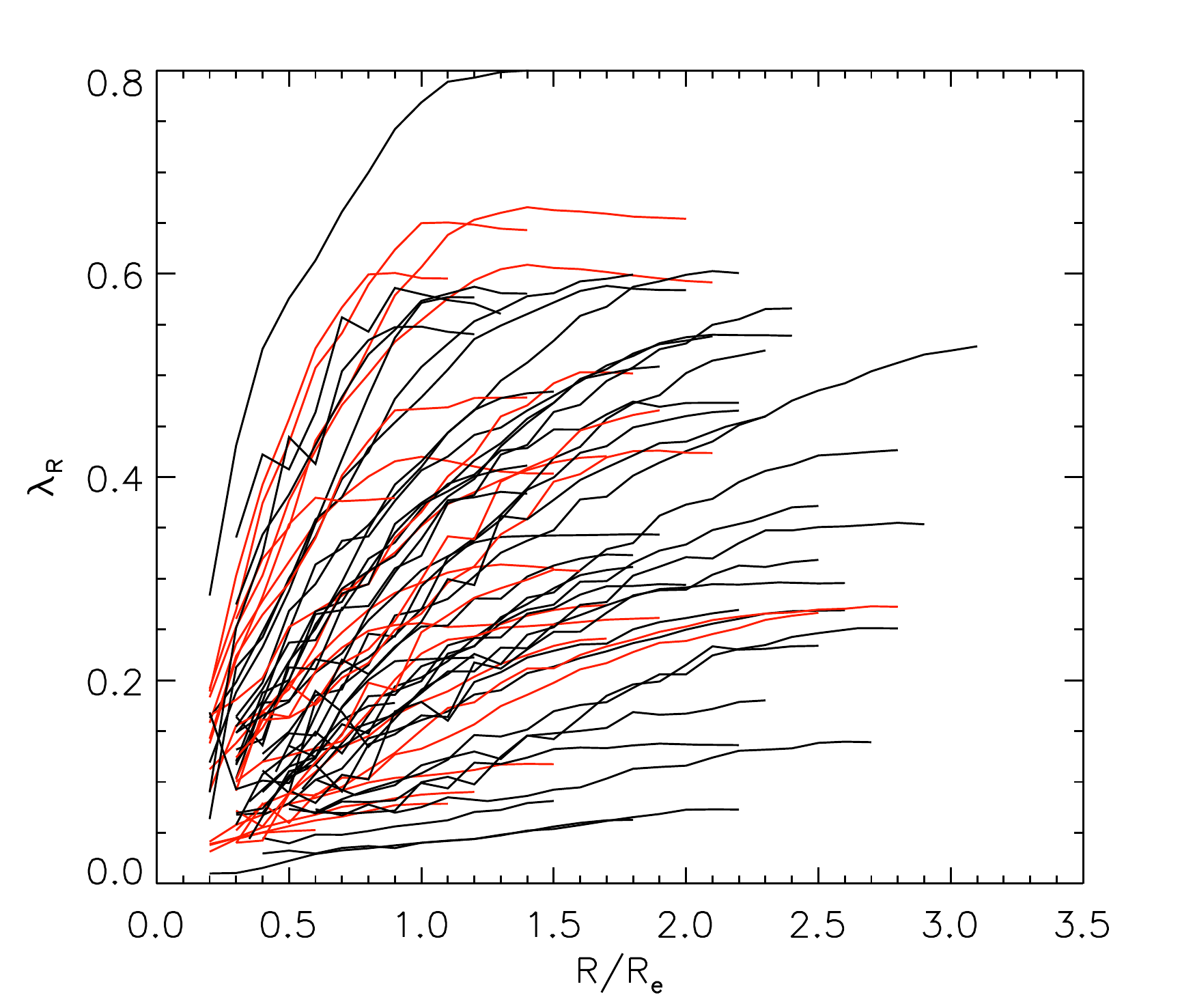}
    \caption
     {$\lambda_{\rm R}$ profile distribution. The black and red lines indicate the distribution of undisturbed and morphologically-disturbed galaxies. A gradual distribution of $\lambda_{\rm R}$ from slow to fast is detected, rather than two distinct classes. }
    \label{lambda}
    \end{figure}

We measured the spin parameter $\lambda_{\rm R}$ integrating within elliptical apertures with fixed $PA$ and $\epsilon$ at the effective radius (see Section 3.2) following Emsellem et al. (2007):
\begin{equation}
\lambda_{\rm R} \equiv {\left\langle R\left| V \right| \right\rangle \over \left\langle R\sqrt{V^2+\sigma^2} \right\rangle} = {{\sum_{i=1}^N F_{i} R_{i}\left| V_{i} \right|}\over{{\sum_{i=1}^N  F_{i} R_{i}\sqrt{V^{2}_{i}+\sigma^{2}_{i}} }}}
\end{equation}
 where $F_i$, $R_i$, $V_i$, and $\sigma_i$ are the flux, radius, velocity, and velocity dispersion of the $i$th spaxel, respectively.
For the galaxies whose isophote at the effective radius is highly affected by inner features, such as bars, we measured $\lambda_{\rm R}$ using the values of $\epsilon$ and $PA$ at $\mu_{\rm r} = 25\,\sur$ which corresponds to a semi-major axis  $3-5\, R_{\rm e}$. Radial profiles of $\lambda_{\rm R}$ are shown in Figure~\ref{lambda}.
Our cluster galaxies do show a wide range of spin parameter reaching high values, as has been reported by previous studies. The red curves show the morphologically-disturbed galaxies; and from a cursory inspection, there does not appear to be a clear distinction between disturbed and undisturbed galaxies.

\begin{figure}[tb]
    \centering
    \includegraphics[width=0.5\textwidth]{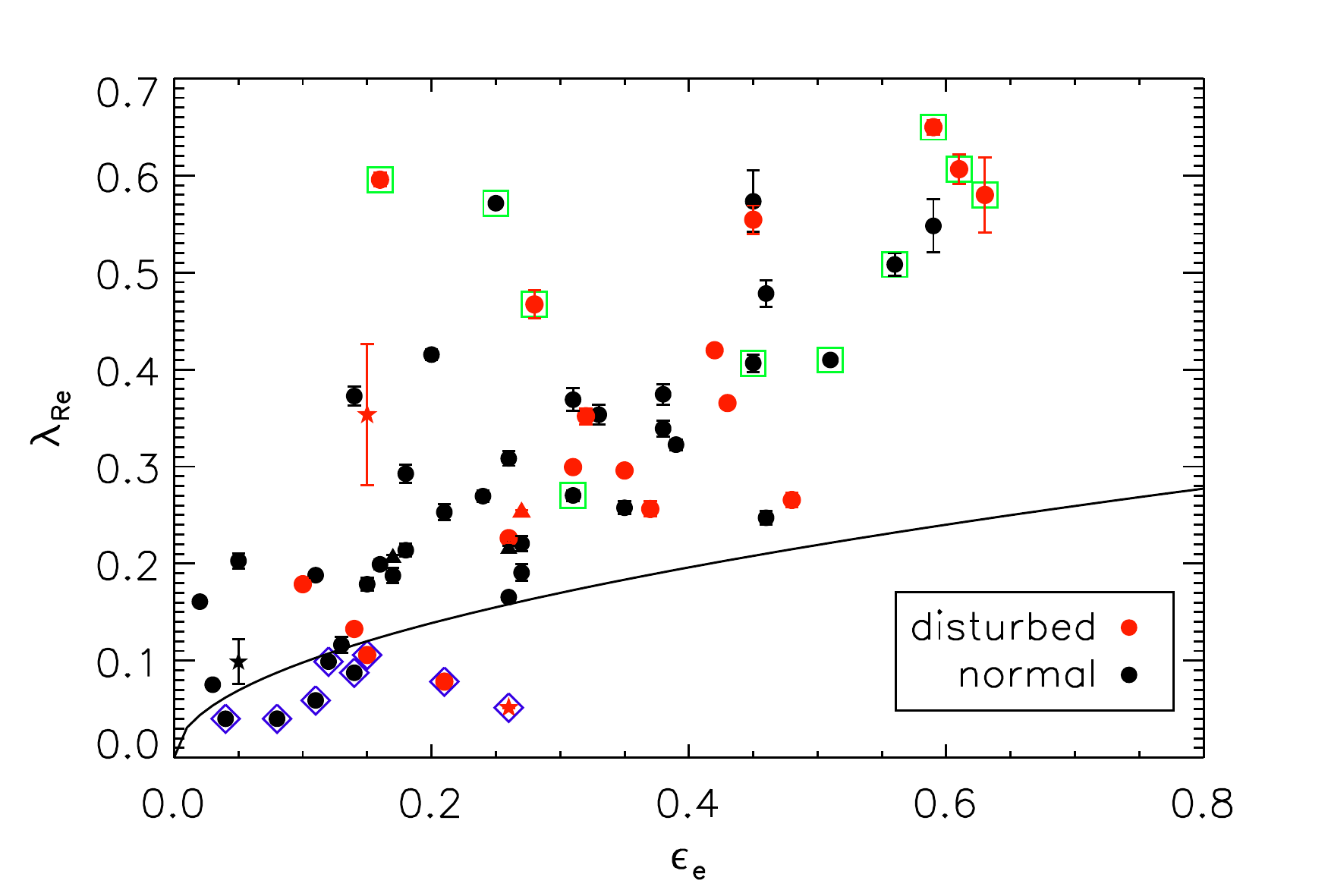}
    \caption[The $\lambda_{\rm R}$ - $\epsilon$ plane]
     {The $\lambda_{\rm R_{\rm e}} - \epsilon$ plane for the diagnostic of slow rotators. The black and red symbols indicate morphologically-disturbed and undisturbed galaxies, respectively. Slow rotators (open diamonds) are defined as galaxies under the demarcation line at $R_{\rm e}$ from Emsellem et al. (2011). We used $\lambda_{\rm R}$ at $0.5\,R_{\rm e}$ (stars) and $2\,R_{\rm e}$ (triangles) for some galaxies. Error bars on $\lambda_{\rm R}$ were calculated from the analytic method described in Houghton et al. (2013). Approximately, 37\% (3/8) of slow rotators show disturbed features. Late-type galaxies (open squares) tend to have high $\lambda_{\rm R_{\rm e}}$. }
    \label{fig_lambda2}
    \end{figure}

We present the result first in the $\lambda_{\rm R} - \epsilon$ plane in Figure~\ref{fig_lambda2}. The demarcation line of
\begin{equation}
\lambda_{\rm R} < k\sqrt{\epsilon} , 
\end{equation}
where $k$ is the coefficient for fiducial radius, has been proposed to select slow and fast rotators by previous studies. We adopt $k$ = 0.265, 0.310, and 0.363 for R = $0.5\,R_{\rm e}$, $R_{\rm e}$, and $2\,R_{\rm e}$ from Emsellem et al. (2011) and Fogarty et al. (2014), respectively.

Considering the radial dependence of kinematics, it is desirable to measure kinematic properties at a consistent radius in all galaxies (Arnold et al. 2014; Foster et al. 2016). We used the diagnostic for the effective radius for all the galaxies in our sample except six. For three galaxies including the BCG, we used the diagnostic for $0.5\,R_{\rm e}$ because the SAMI map did not reach out to $R_{\rm e}$. This is not a problem because the two non-BCG galaxies are already classified as fast rotators at $0.5\,R_{\rm e}$, and so if their $\lambda_{\rm R}$ rises with radius as in most cases, they would still be classified as fast rotators at $R_{\rm e}$. The BCG on the other hand is classified as a slow rotator at $0.5\,R_{\rm e}$, and given its radial profile would most likely be a slow rotator at larger radii as well. For the other three galaxies whose $R_{\rm e} < 1.5$ arcsecs, we used the diagnostic for $2\,R_{\rm e}$ instead in order to avoid the beam smearing effect. The seeing effect on $\lambda_{\rm R_{\rm e}}$ measurement is discussed in van de Sande et al. (in prep.).

Out of the 63 galaxies in our sample, 8 galaxies were classified as slow rotators (13\%). This fraction increases to 15\% if the sample is limited to early-type galaxies (53 galaxies) and is consistent with previous studies that found slow rotators in cluster environments using the same classification as this study (Cappellari et al. 2011; D$'$Eugenio et al. 2013; Houghton et al. 2013; Fogarty et al. 2014; Scott et al. 2014). Of the eight slow rotators three galaxies exhibited disturbed features, which is comparable to previous reports (Duc et al. 2011; Jimmy et al. 2013). We found a slightly higher fraction of slow rotators in morphologically-disturbed galaxies ($15\pm10\%$) than in their undisturbed counterparts ($12\pm 6\%$).

The effective rotational velocity ($V_{\rm e}$) is defined as the modeled amplitude of the rotational velocity ($k_1$) at $R_{\rm e}$ as measured from $kinemetry$. SAMI could not recover full rotation curves for most of our sample because of the limited FOV (15$''$) and low $S/N$ in outer spaxels. Therefore, we could not estimate the maximum flattened velocity for the whole sample. Instead, we measured the rotational velocity at $R_{\rm e}$. The $V_{\rm e}$ for the three galaxies whose SAMI measurements did not reach out to $R_{\rm e}$ was estimated by the linear extrapolation using the closest 3 points of $k_1$. 
We confirm that there is a linear correlation between $V_{\rm e}$ and the maximum rotational velocity based on 20 galaxies whose velocity profile reaches out to a flattened velocity curve.

The $V_{\rm e}$ was corrected for inclination using the apparent major ($a$) and minor ($b$) axes at $\mu_{\rm r}$ = 25 $\sur$ using the following formula:
 \begin{equation}
 cos (i) = \sqrt {{(b/a)^2 - q_0^2} \over {1-q_0^2}}
 \end{equation}
where intrinsic shape parameter, $q_0$, is set to be 0.65 for early-type and 0.2 for late-type galaxies, following Catinella et al. (2012). Although the inclination correction is highly uncertain especially for early-type galaxies, it has a negligible effect on our results and conclusions.

\begin{figure}[tb]
    \centering
    \includegraphics[width=0.5\textwidth]{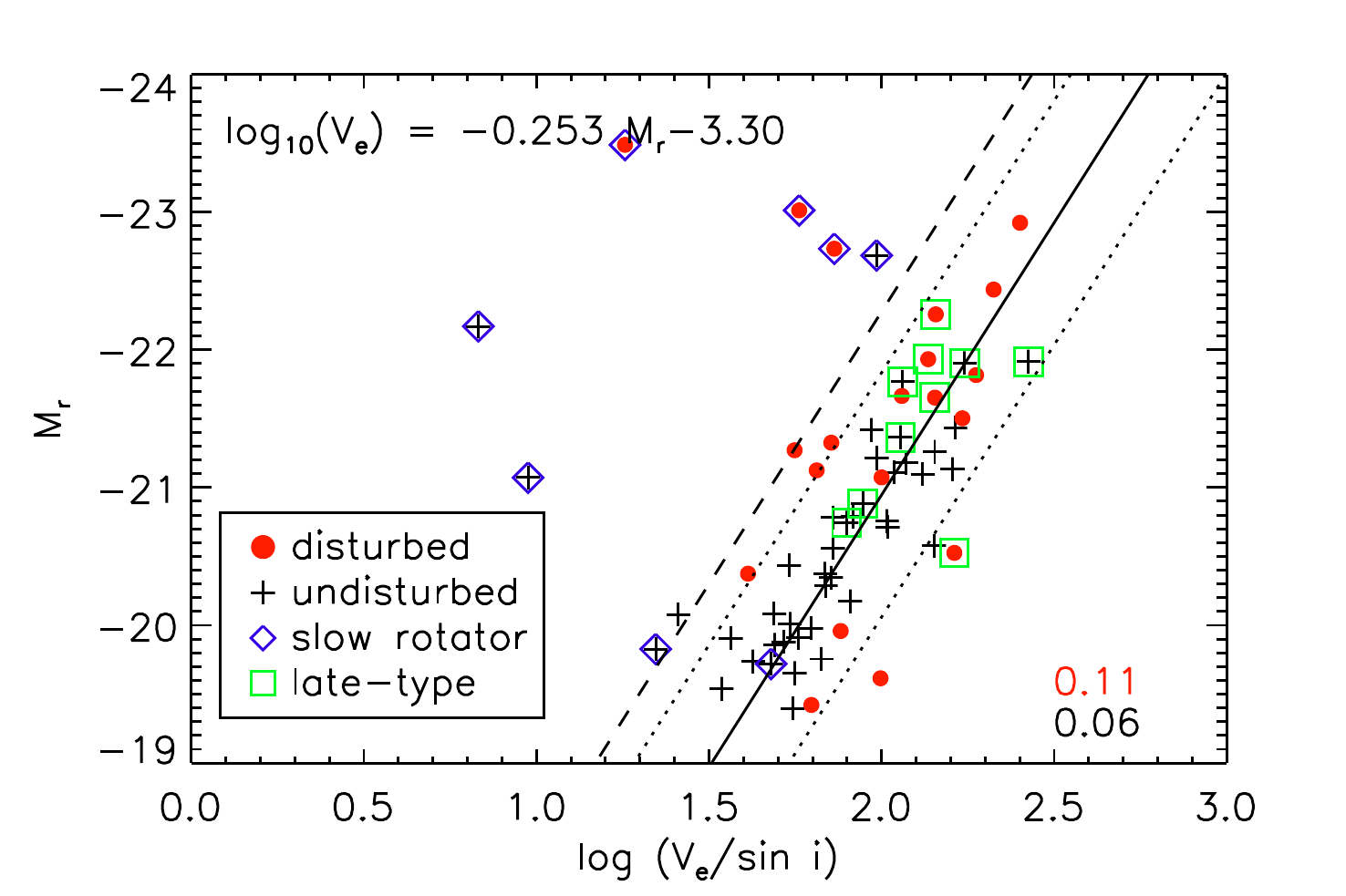}
    \caption[The T-F relation]
     {The effective T-F relation. Morphologically-disturbed and undisturbed galaxies are denoted by the red and black symbols, respectively. The solid line is the inverse fit of the undisturbed galaxies using the $M_{\rm r}$ as an independent variable (black symbols). The dotted and dashed lines correspond to 2$\sigma$ and 3$\sigma$ of the relation for undisturbed galaxies, respectively. Slow rotators defined using the $\lambda_{\rm R} - \epsilon$ plane are denoted by open diamonds (See Figure~\ref{fig_lambda2}), and they deviate significantly from the T-F relation of undisturbed galaxies. Scatter from the least-squares fit was calculated on the morphologically-disturbed (0.11) and undisturbed galaxies (0.06).}
    \label{fig_tully}
    \end{figure}

We present the ``effective'' T-F relation using $V_{\rm e}$ in Figure~\ref{fig_tully}. Our strategic use of velocities from kinemetric modeling significantly reduces the velocity scatter and restores the T-F relation, even for early-type galaxies which have generally been known to lie away from the sequence. As a result, most of the galaxies in our sample are well aligned with the relation. The few outliers seen in Figure~\ref{fig_tully} are very slow for their stellar mass and most of them (7 out of 8) are classified as slow rotators in the $\lambda_{\rm R}-\epsilon$ plane (Figure~\ref{fig_lambda2}). One slow rotator (cross inside diamond at ${\rm log}\,(V_{\rm e}/{\rm sin}\, i) = 1.7$ and $M_{\rm r}=-19.6$) lies closely to the T-F relation, possibly due to an inaccurate inclination correction. The undisturbed galaxy that lies beyond the 3$\sigma$ (dashed) line (cross at  ${\rm log}\,(V_{\rm e}/{\rm sin}\, i) = 1.4$ and $M_{\rm r}=-20$) on the other hand, likely suffered from the beam smearing effect and hence the inaccurate estimate of $V_{\rm e}$.
 
We estimated the scatter by first finding the robust least-squares fit (the solid line) to the unperturbed galaxies and then measuring the standard deviation of residuals in velocity from the fit excluding slow rotators. Our T-F relation based on the {\em derived} values of $V_{\rm e}$ shows scatter that is comparable to that of late-type galaxies (see Kannappan et al. 2002).
We looked for a relation between the stellar velocity dispersion at $R_{\rm e}$ ($\sigma_{\rm e}$) and the scatter of the T-F relation and found that only slow rotators, which have high velocity dispersion, also have large scatter (over $3\sigma$ from the T-F relation). Fast rotators, regardless of if they are morphologically disturbed or not, typically have a low scatter ( $\le 3\sigma$) from the T-F relation, but show a range of $\sigma_{\rm e}$. The possible physical reasons for early-type galaxies lying on the T-F relation will be investigated in a future paper.

 Morphologically-disturbed galaxies show a larger scatter (0.11 dex) in the T-F relation relative to the undisturbed galaxies (0.06 dex). Kassin et al. (2007) reported that perturbed $disk$ galaxies deviate from the T-F relation by always showing slower rotation, whereas we found that morphologically-disturbed galaxies scatter in both directions in velocity. Flores et al. (2006) and Puech et al. (2010) reported large scatter on the T-F relation in galaxies with complex kinematics or perturbed rotation, which probably relates to galaxy mergers. Galaxy mergers generally enhance luminosity but may either reduce or enhance their rotation speed depending on the merger/interaction geometry. 

\subsection{Misalignment between photometric $PA$ and kinematic orientation}
\label{sec:pa}
    \begin{figure}[tb]
    \centering
    \includegraphics[width=0.5\textwidth]{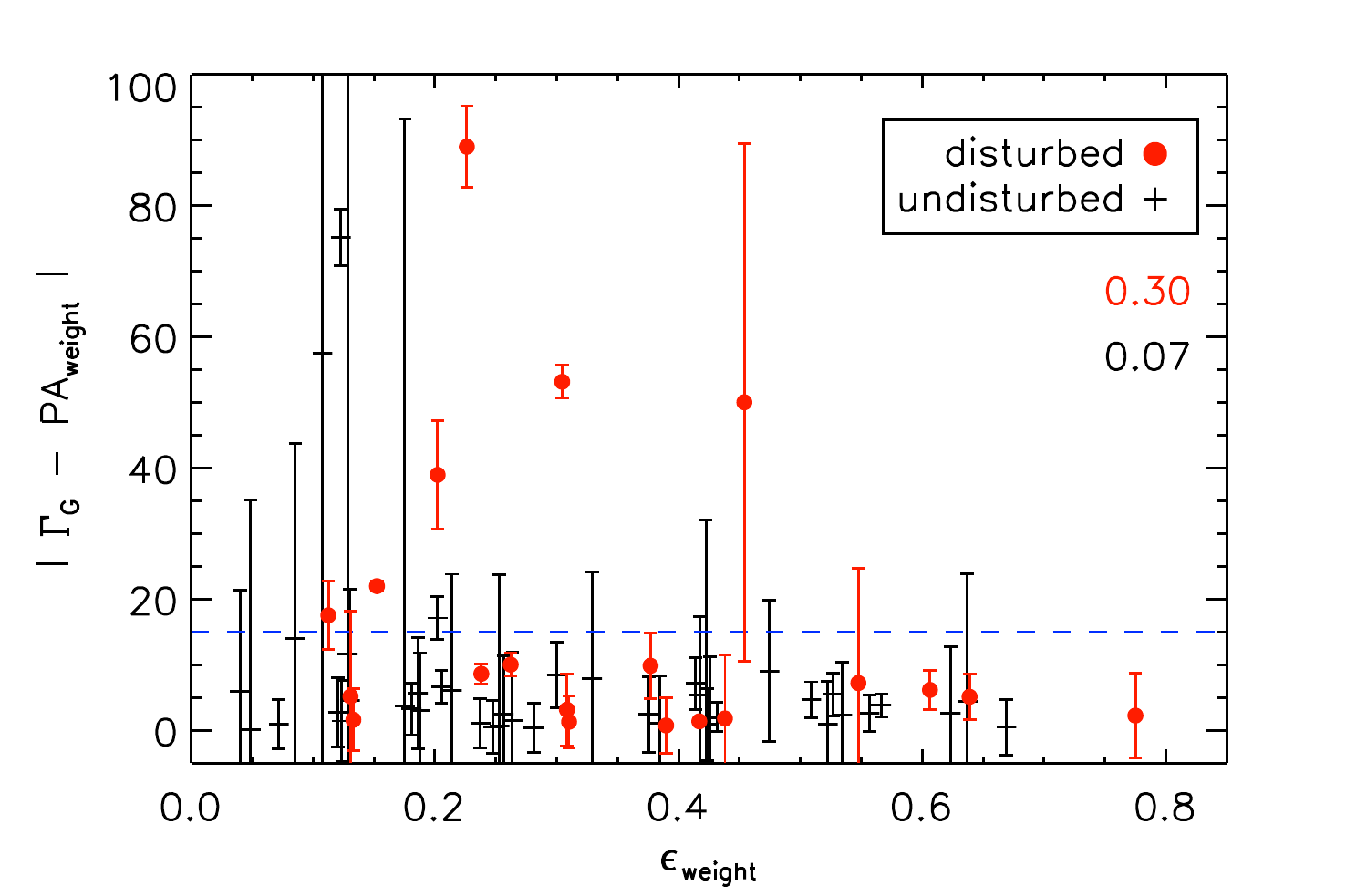}
    \caption[Kinematic misalignment]
     {Kinematic misalignment as a function of $\epsilon_{\rm weight}$. Morphologically-disturbed and undisturbed galaxies are denoted by the red and black symbols, respectively. The error bars indicate a $3\sigma$ confidence level on $\Gamma_{\rm G}$. The dashed line corresponds to $15\deg$ dividing angle of kinematic misalignment. Fractions of galaxies showing the kinematic misalignment ($>15\deg$) in disturbed (red) and undisturbed (black) galaxies are shown in the top-right corner. Morphologically-disturbed galaxies show 4 times higher fraction of kinematic misalignment, and most of misaligned galaxies displayed merger signatures.}
    \label{misalign1}
    \end{figure}

Some galaxies with disturbed features show misalignment in orientation between the photometry and kinematics. The global kinematic orientation ($\Gamma_{\rm G}$), indicating the mean motion of the stellar velocity, was measured using the method described in Krajnovi\'c et al. (2006). Specifically, we made 361 simulated maps by rotating the original map in steps of 0.5$\deg$ and took the angle that minimized $\chi^2$. The uncertainty in $\Gamma_{\rm G}$ is defined as the minimum opening angle of simulated maps that satisfied a $3\sigma$ confidence level; $\Delta \chi^2 < 9 + 3 \sqrt{2 N}$ (van den Bosch \& van de Ven 2009; Krajnovi\'c et al. 2011). The mean uncertainty value on $\Gamma_{\rm G}$ is 6$\deg$ for the sample galaxies; so, we used $15\deg$ as a cut for ``misaligned'' galaxies. 
We measured a luminosity-weighted average of position angle ($PA_{\rm weight}$) and ellipticity ($\epsilon_{\rm weight}$) at $R \geq 2\,R_{\rm e}$  as representative parameters in photometry.

There are several uncertainties on the measurement of $PA$ in both photometry and kinematics. 
First, $PA$ cannot be determined in round galaxies. The central regions of galaxies ($R \lesssim R_{\rm e}$) are rather circular, and therefore $PA$ measurements at $R_{\rm e}$ from optical images have large uncertainties. We therefore measured $PA$ in outer regions ($R \geq 2\,R_{\rm e}$) as routinely exercised (e.g. Cappellari et al. 2007, Krajnovi$\acute{\rm c}$ et al. 2011). 
Second, photometric $PA$ at specific radius can follow the dominant light from spiral arms or disturbed features. A luminosity-weighted average of $PA$ in this study is free from this issue. We manually confirmed that $PA_{\rm weight}$ does not follow local features.
Third, we cannot determine the kinematic orientation for galaxies with no rotation. However, even slow rotators in this study have a certain level of rotation which is enough to judge their kinematic orientation, except two galaxies whose $V_{\rm e}$ is below 10 km/s. We confirmed that our results are not affected by excluding them from this analysis.


Galaxy interactions seem to have played a key role in creating the misalignment between $\Gamma_{\rm G}$ and $PA_{\rm weight}$ (Figure~\ref{misalign1}). Galaxies with disturbed features show a 4 times larger fraction ($0.3\pm0.16$) of misaligned galaxies ($\geq15\deg$) compared to their undisturbed counterparts ($0.07\pm0.06$). We could not find significant differences in the misalignment even if we measure $PA_{\rm weight}$ at $R \geq R_{\rm e}$ or $R \geq 3\,R_{\rm e}$. Krajnovi$\acute{\rm c}$ et al. (2011) investigated the kinematic misalignment with regards to galaxy morphology and found a higher fraction of kinematic misalignment greater than $15\deg$ in their ``interaction'' or ``shell'' class (24\%) than in their normal morphological class (9\%). Barrera-Ballesteros et al. (2015) also reported that 43\% of the ``interacting'' sample showed misalignment larger than the mean value of misalignment angle of the control sample.

Based on the measurement at $R \geq 2 \,R_{\rm e}$, we found that 67\% (6/9) of misaligned galaxies showed disturbed features. In Krajnovi$\acute{\rm c}$ et al. (2011), however, only 18\% of misaligned galaxies were classified as ``interaction'' or ``shell'' class. Our images are three magnitudes deeper than those used by Krajnovi$\acute{\rm c}$ et al. (2011) (from the SDSS and Isaac Newton Telescope), and thus reveal the link between kinematic misalignment and galaxy interaction more clearly.

\begin{figure}[tb]
    \centering
    \includegraphics[width=0.5\textwidth]{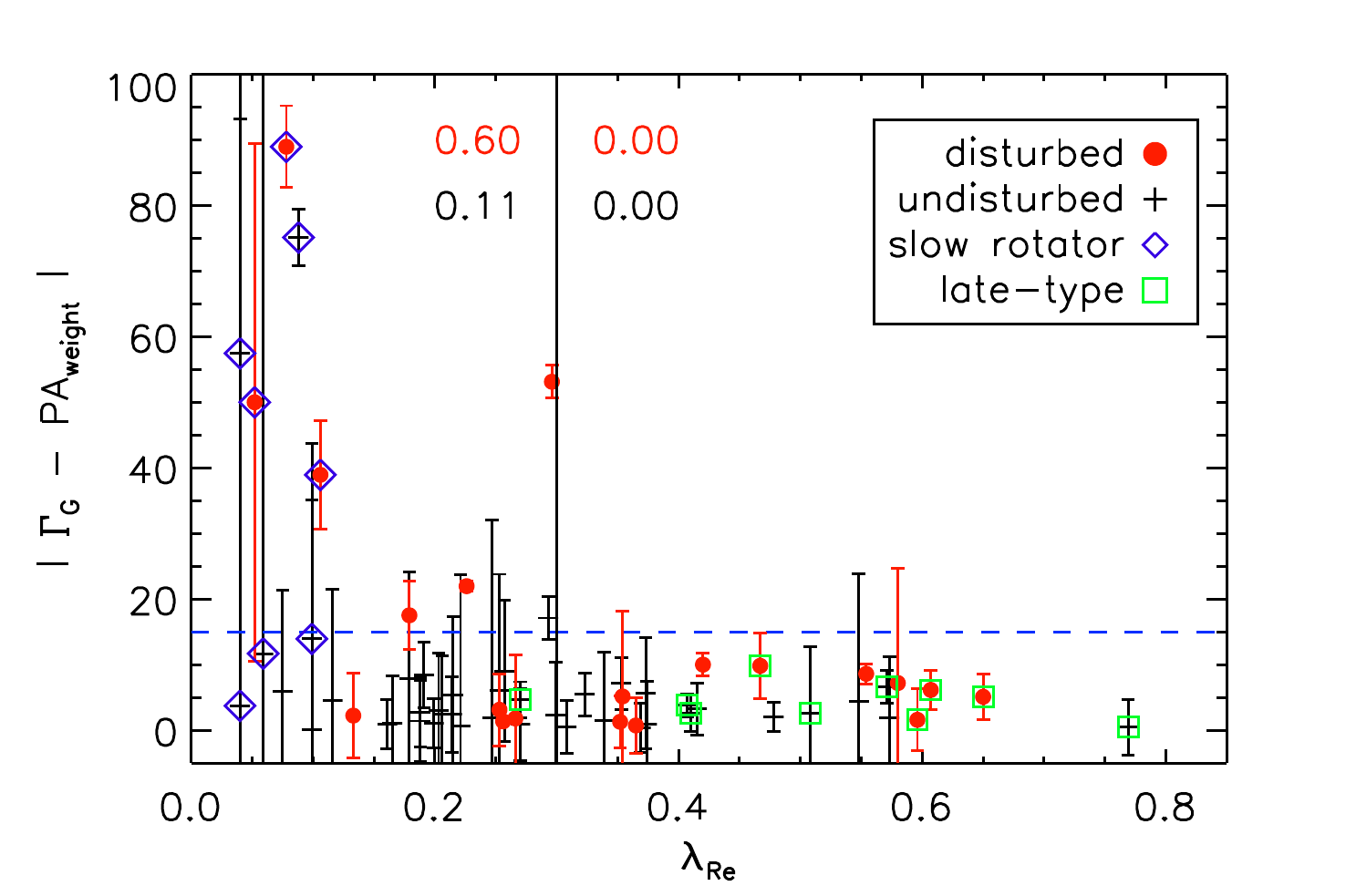}
    \caption[Kinematic misalignment and $\lambda_{\rm R_{\rm e}}$]
     {Kinematic misalignment plotted as a function of $\lambda_{\rm R_{\rm e}}$. The dashed line indicates $|\Gamma_{\rm G} - PA_{weight}| = 15\deg$, the dividing angle of the kinematic misalignments in this study. The error bars indicate a $3\sigma$ confidence level on $\Gamma_{\rm G}$. Kinematic misalignment is only seen in galaxies having low angular momentum (e.g. $\lambda_{\rm R_{\rm e}} <  0.3$); over half of the morphologically-disturbed galaxies (0.60) with low angular momentum are misaligned. Undisturbed galaxies show small misalignment fraction (0.11) even when the spin parameter proxy $\lambda_{\rm R_{\rm e}}$ is small.}
    \label{fig_misalign2}
    \end{figure}

Misalignment has been reported to be more conspicuous in slow rotators (e.g., Cappellari et al. 2007, Krajnovi$\acute{\rm c}$ et al. 2011). In Figure~\ref{fig_misalign2} we also observe a similar trend. We found that only four out of the 55 fast rotators are misaligned, compared to five out of the eight slow rotators. In addition, we found that  kinematic misalignment is only seen in galaxies with low angular momentum (e.g. $\lambda_{\rm R_{\rm e}} < 0.3$). 
If we only consider galaxies with $\lambda_{\rm R_{\rm e}} < 0.3$, a contrast between the morphologically-disturbed and undisturbed galaxies becomes more significant: undisturbed galaxies still have a small fraction (0.11) of kinematic misalignment; whereas 60\% of the morphologically-disturbed galaxies show misalignment. Therefore, we can conclude that mergers   seem to affect the kinematic orientation of the galaxies primarily with low angular momentum.

\subsection{Kinematic asymmetry}
\label{sec:say}

    \begin{figure}[tb]
    \centering
    \includegraphics[width=0.5\textwidth]{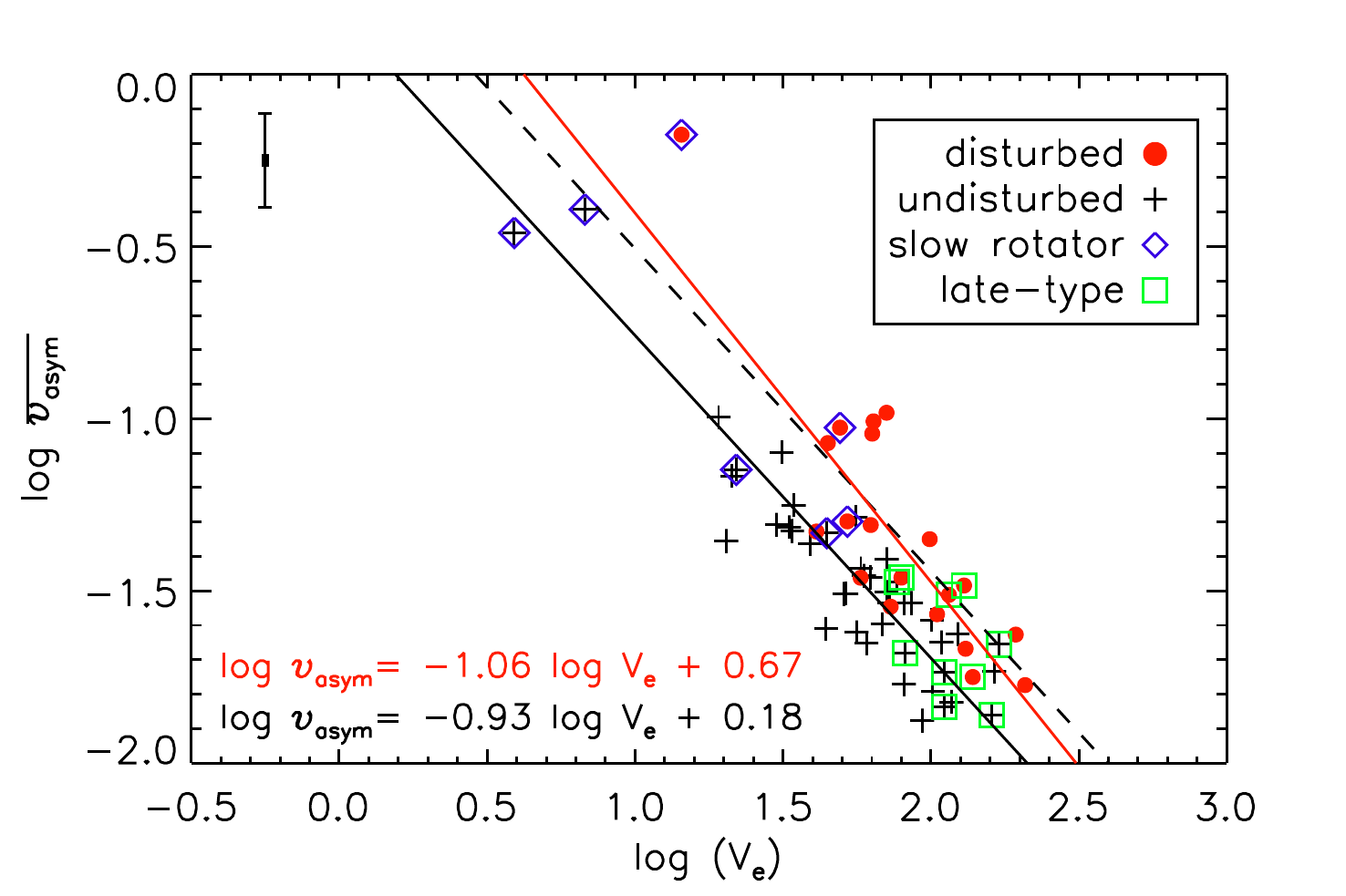}
    \caption[Kinematic asymmetry]
     {Kinematic asymmetry $\overline{\upsilon_{\rm asym}}$ has a tight correlation with rotational velocity $V_{\rm e}$. Inclination correction was not applied in this figure. The solid- and dashed-black lines show the least-squares fit and 2$\sigma$ distribution of undisturbed galaxies ($\sigma=0.125$ dex in $\overline{\upsilon_{\rm asym}}$), respectively. The solid-red line indicates the least-squares fit of morphologically-disturbed galaxies. We present the fits of disturbed (red) and undisturbed (black) galaxies in the bottom left, and the typical errors are shown in the top left. In general, kinematic asymmetry in morphologically-disturbed galaxies is boosted compared to that in undisturbed galaxies of the same $V_{\rm e}$.}
    \label{fig_vasym}
    \end{figure}

 We measured the kinematic asymmetry in the stellar velocity field ($\upsilon_{\rm asym}$) using the first ($k_1$) coefficient and the summation of the odd high order ($k_3+k_5$) coefficients from $kinemetry$ following Bloom et al. (2016, submitted): 
\begin{equation}
\upsilon_{\rm asym} = {k_3+ k_5 \over 2k_1}.
\end{equation}
This $\upsilon_{\rm asym}$, using only the odd terms, is a modification of kinematic asymmetry used by Shapiro et al. (2008), who defined kinematic asymmetry using both odd and even terms ($k_2,k_3,k_4$, and $k_5$). We took the mean value of $\upsilon_{\rm asym}$ at each spaxel without giving luminosity weights because our study is focused on faint features.

Previous studies that parameterized kinematic asymmetry used specific values of kinematic asymmetry, e.g., $k_5/k_1$ or $(k_3+k_5)/2k_1$, to identify ``non-regular'' kinematics (e.g. Krajnovi$\acute{\rm c}$ et al. 2011, Bloom et al. 2016, submitted); however, a further consideration of rotational velocity might be necessary. 
The $\overline{\upsilon_{\rm asym}}$ has a tight correlation with $V_{\rm e}$, $k_1$ measured at the effective radius, due to its definition including $k_1$ (Equation (6)). Therefore, galaxies with small $V_{\rm e}$ show large $\overline{\upsilon_{\rm asym}}$, spontaneously (Figure 8).


Morphologically-disturbed galaxies show a large $\overline{\upsilon_{\rm asym}}$ compared with undisturbed galaxies at a given $V_{\rm e}$. The least-squares fit to the morphologically-disturbed sample (the solid-red line) is roughly parallel to and 2$\sigma$ away from the sequence for the undisturbed sample (the solid-black line), which indicates that galaxy interactions generally increase $\overline{\upsilon_{\rm asym}}$.

Krajnovi$\acute{\rm c}$ et al. (2011) reported that only a small fraction of galaxies show signatures of past interactions {\em regardless} of the presence of kinematic anomaly based on the simple cut of $k_5/k_1$ = 0.04.  Following a similar horizontal cut in Figure 8, we would find the same result. However, our sample spans a larger baseline in $V_e$  and clearly calls for necessity for considering a $V_e$-dependent cut rather than a uniform cut to conclude on the impact of dynamical interactions on kinematic anomaly. We discuss this in the next section.


\section{Discussion}
\label{sec:discussion}

\subsection{Effect of galaxy interactions on kinematics}
 We now discuss how kinematic anomalies relate to each other. First, we define two parameters to quantify scatter in the T-F relation ($S_{\rm TF}$) and the perturbations in $\overline{\upsilon_{\rm asym}}$ ($P_{\rm asym}$):
 \begin{eqnarray}
 S_{\rm TF} = \left|{{\rm log}\,{V_{\rm e} \over sin\,i}} - {\rm log}\,V_{\rm e,fit}\right|  \\
  P_{\rm asym} = {\rm log}\,\overline{\upsilon_{\rm asym}}- {\rm log}\,\overline{\upsilon_{\rm asym,fit}}
 \end{eqnarray}
 where log$\,V_{\rm e,fit}$ is the fitted rotational velocity for a given luminosity (the solid line in Figure~\ref{fig_tully}), and ${\rm log}\,\upsilon_{\rm asym,fit}$ is the fitted kinematic asymmetry for a given ${\rm log}\,V_{\rm e}$ (the solid-black line in Figure~\ref{fig_vasym}). Therefore, $S_{\rm TF}$ defines scatter in the rotational velocity for a given luminosity. We use the absolute value because mergers appear to cause scatter in both directions. $P_{\rm asym}$ is the excess of $\rm log\,\overline{\upsilon_{\rm asym}}$ compared with the fitted value of undisturbed galaxies for a given rotation.
 
\begin{figure}[tb]
    \centering
    \includegraphics[width=0.5\textwidth]{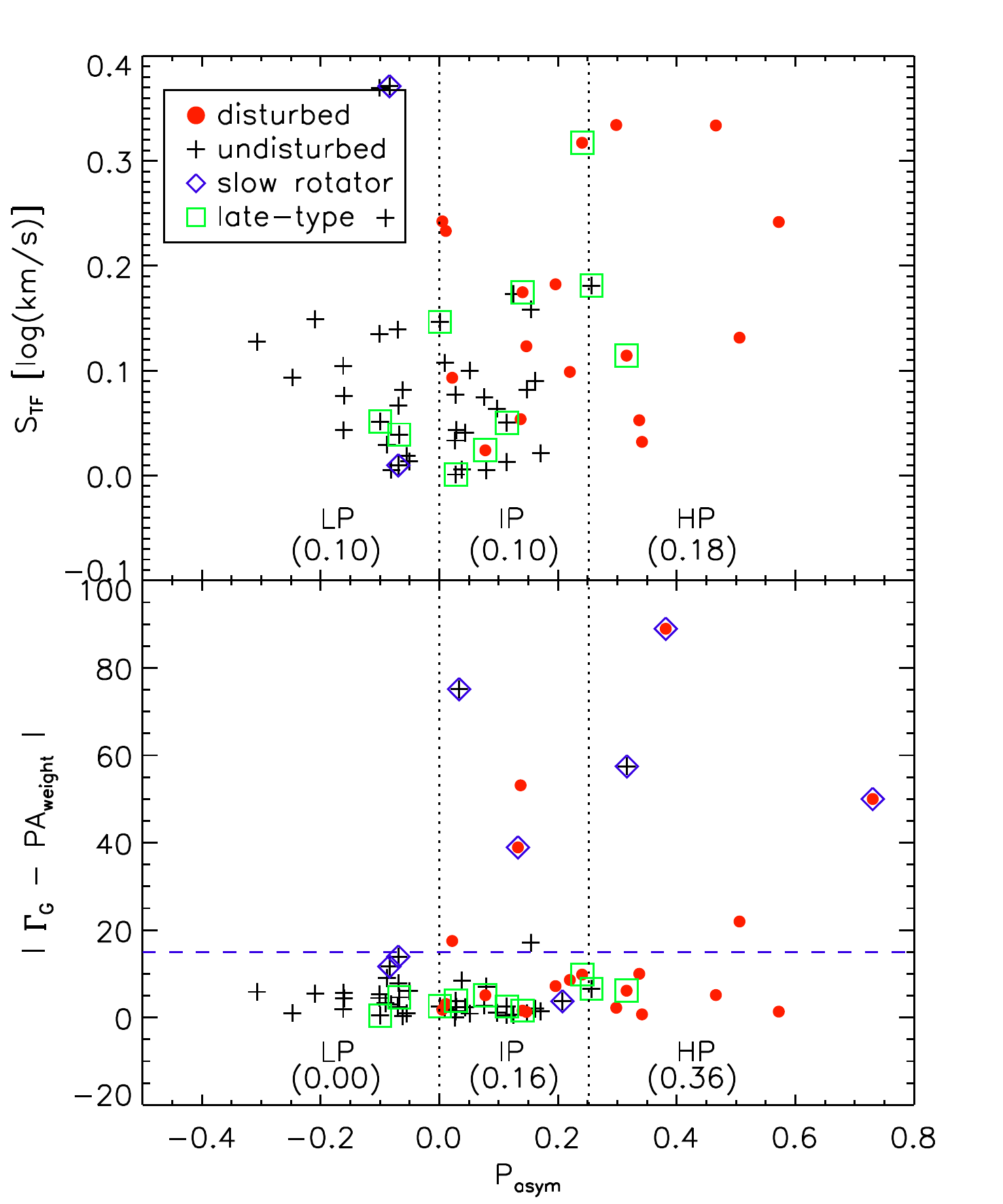}
    \caption[Correlations of kinematic anomalies]
     {Correlations of the kinematic anomalies in orientation ($|\Gamma_{\rm G} - PA_{\rm weight}|$), kinematic asymmetry ($P_{\rm asym}$), and angular momentum ($S_{\rm TF}$). Morphologically-disturbed and undisturbed galaxies are denoted by the red and black symbols, respectively. Late-type galaxies and slow rotators are indicated by open rectangles and diamonds, respectively.
     In the top panel, most of the slow rotators are outside the scope of the y-axis. Galaxies are divided into 3 groups (LP, IP, and HP) according to their $P_{\rm asym}$. The mean value of $S_{\rm TF}$ (top panel) or fractions of misalignment (bottom panel; $|\Gamma_{\rm G} - PA_{\rm weight}|>15\deg$) for each group is provided in the parentheses. The HP group shows a larger mean value of $S_{\rm TF}$ and a higher fraction of misalignment than the others.} 
    \label{fig_resi}
    \end{figure}

 We present a comparison between three kinematic parameters  in Figure~\ref{fig_resi}. 
 For easy comparison, we divided galaxies into three groups according to $P_{\rm asym}$: low levels of perturbation (LP; $P_{\rm asym}<$ 0; the least-squares fit of undisturbed galaxies), intermediate perturbation (IP; $0\leq P_{\rm asym}\leq 0.25$), and highly perturbed kinematics (HP; $P_{\rm asym} > 0.25$; 2$\sigma$ of the undisturbed galaxies).
 Except for slow rotators, galaxies with highly perturbed kinematics (HP) show a higher mean value of $S_{\rm TF}$ ($0.18\pm 0.12$) than galaxies with intermediate or low levels of perturbation (IP or LP) whose mean $S_{\rm TF}$ is $0.10\pm 0.08$, although scatter is too large to make this trend statistically significant. 
The trend, if real, is consistent with previous studies based on visual classification on kinematic morphology (Flores et al. 2006; Puech et al. 2010). 

The bottom panel of Figure 9 shows a comparison between the kinematic misalignment ($|\Gamma_{\rm G} - PA_{\rm weight}|$) and perturbation in $\overline{\upsilon_{\rm asym}}$ ($P_{\rm asym}$).
Kinematic misalignment ($|\Gamma_{\rm G} - PA_{\rm weight}|>15\deg$) is only shown in IP and HP group, and we found a higher fraction of misalignment in HP (0.36) than in IP (0.16).
 
The mild yet positive correlations between the three parameters quantifying kinematic anomalies suggest that galaxy interactions and mergers have a key role in causing such asymmetries.

\subsection{Asymmetry}

So far we have relied primarily on visual selection of mergers. This is because we believe that human eyes detect morphological anomalies effectively and simply enough. As a sanity check, we now measure the degree of disturbed features using the CAS asymmetry parameter ($A$) on photometric data (Abraham et al. 1996; Conselice 2003).

 Following the classic approach, we subtracted the 180$\deg$ rotated image from the original one and normalized it using the total flux. Then, the asymmetry parameter $A$ is measured by the summation of residuals excluding contaminated regions by other objects. The central parts of bright galaxies are saturated in our deep images because we were attempting to detect low surface brightness features; hence, we only considered residuals in the outer regions: beyond $R_{\rm e}$ to $3\sigma$ above the sky level. For this reason, there is a need for caution when making direct comparisons against previous studies employing asymmetry parameters.

\begin{figure}[tb]
    \centering
    \includegraphics[width=0.5\textwidth]{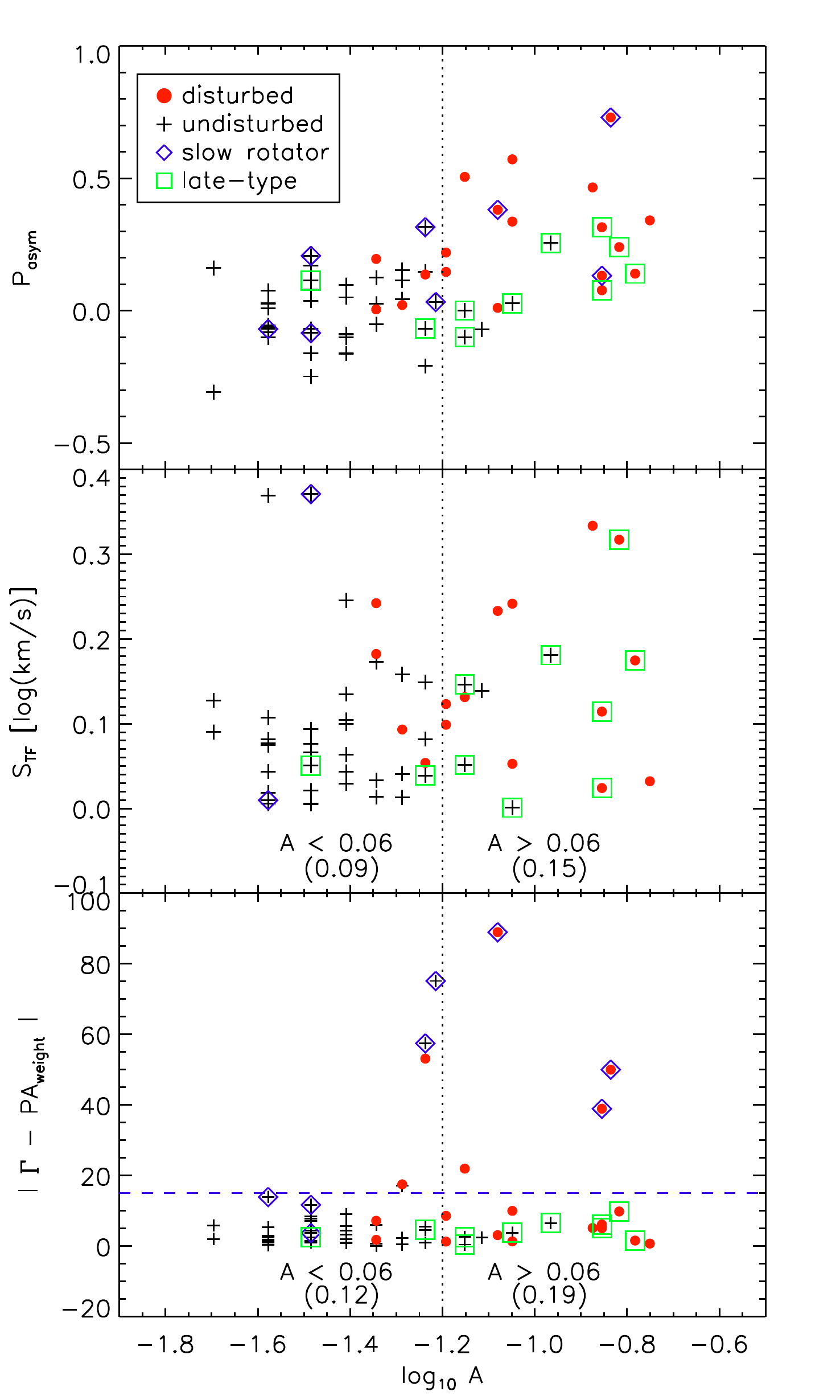}
    \caption[Photometric asymmetry]
     {A comparison between the kinematic anomalies and the photometric asymmetry $A$. Morphologically-disturbed and undisturbed galaxies are denoted by the red and black symbols, respectively. Slow rotators are denoted by open diamonds, and late-type galaxies are indicated by open rectangles. The photometric asymmetry $A$ and the perturbation on stellar kinematics ($P_{\rm asym}$) are correlated to each other (top panel). The mean value of $S_{\rm TF}$ (middle panel) or fractions of misalignment (bottom panel; $|\Gamma_{\rm G} - PA_{\rm weight}|>15\deg$) for each group is provided in the parentheses. Galaxies with $A > 0.06$ shows a larger mean value of $S_{\rm TF}$ and a higher fraction of misalignment than galaxies with  $A < 0.06$. }
    \label{fig_asym}
    \end{figure}

 In this study, even the morphologically-disturbed galaxies have $A<0.2$; i.e., we are indeed dealing with low surface brightness features that are only detectable in deep images (Figure~\ref{fig_asym}). 
Our visually selected disturbed galaxies were well distinguished by $A$, except the late-type galaxies, which naturally have a high value of $A$ due to spiral arms. 
Over 80\% of the morphologically-disturbed galaxies have $A > 0.06$ as shown by the vertical line in Figure~\ref{fig_asym}.
In fact, apart from late-type galaxies (green squares), all but one galaxies in the high value of $A$ ($>0.06$) region are morphologically disturbed.

Conselice, Rajgor, \& Myers (2008) pointed out that it is difficult to identify visually-selected  disturbed galaxies using asymmetry parameters if $A \lesssim 0.35$. However, we found a good separation between disturbed and undisturbed galaxies based on low surface brightness features measured in $A$. The ability of $A$ to detect peculiarities seems to depend greatly on the depth of the image.

In the top panel, the kinematic and photometric asymmetries seem to be correlated with each other, even for slow rotators and late-type galaxies. However, we could not find a good correlation between photometric asymmetry and the scatter in the T-F relation, $S_{\rm TF}$ (middle panel). We found a higher mean value of $S_{\rm TF}$ in galaxies with $A > 0.06$ ($0.15\pm0.10$), than in galaxies with $A < 0.06$ ($0.09\pm0.08$). 
Kinematic misalignments ($|\Gamma_{\rm G} - PA_{\rm weight}|>15\deg$) appear more in galaxies with $A>0.06$: 19\% (4/21) of galaxies with $A > 0.06$ show misalignment, and 12\% (5/42) of galaxies with  $A < 0.06$ are misaligned (bottom panel). 

The results based on asymmetry index measurements are largely consistent with the result based on the visual inspection presented in the earlier sections.
  
\subsection{Recent mergers and galaxy spin}

Repeated mergers and interactions between galaxies probably reduce the specific angular moment of a galaxy, and hence it is reasonable to expect more frequent merger-related features in slow rotators. 
Our deep images are capable of detecting even (reasonably important) minor mergers and faint features in outer regions of galaxies where dynamical time scale is much longer than in galaxy center and therefore we can probe the signatures of past mergers further back.
We focus on the fraction of disturbed galaxies among slow rotators. 
Three out of 8 slow rotators appear disturbed, which is only marginally significant compared to the fast rotators (17 out of 55). But, the statistical differences do not appear clear enough.

Recently, Duc et al. (2015) investigated galaxy morphology using similar deep images with ours and kinematics of ATLAS$^{\rm 3D}$ sample in the field environment. 
Of the 78 galaxies whose morphological class is not ``Undetermined'', 46 galaxies ($59\%$) displayed features involved in galaxy interactions (``Major merger'' or ``Minor merger'' or ``Interacting'' in their classification). Fractions of galaxy interactions in fast and slow rotators were $58\pm10\%$ (38/66) and $66\pm22\%$ (8/12), respectively in their study. The major difference between their and our results is that they found around 1.8 times higher fractions of galaxy interactions than us. This can be explained by the difference in a sample; their sample includes only galaxies in the field environment. Previous studies using deep images reported that galaxies with post-merger features are more frequently observed in the field (e.g. van Dokkum 2005) than in the cluster environment (e.g. Sheen et al. 2012). Besides, the median distance of their sample is 28 Mpc, which is 7 times closer than our Abell 119 sample. This makes it easier to detect features of morphological disturbance.

Apart from that, slow rotators show only slightly higher fractions of galaxy interactions than fast rotators both in Duc et al. (2015) and in this work. This is perhaps because mergers may have occurred in the distant past on those slow rotators that do not show disturbed features. This is consistent with recent galaxy formation models where important mergers are rare after $z=1$ (Khim et al. 2015).

 As shown in Section 4.1, a single merger (or interaction) can either reduce or enhance the spin depending on the merger condition; that is, both effects are possible, as also demonstrated by simulations (Bois et al. 2011; Naab et al. 2014). This makes it difficult to uniquely determine the effect of mergers on the angular momentum of a galaxy at a specific epoch. 

\section{Summary and prospect}

We investigated the impact of galaxy interactions on the kinematic anomalies using 63 galaxies in Abell 119 based on stellar kinematics from SAMI. We visually selected 20 galaxies with disturbed features ($32\%$) using our deep images. Morphologically-disturbed galaxies in this study have faint features that generally show a slight asymmetry ($A < 0.2$), which can only be detected in deep images.

The use of ``model'' values of rotational velocities reveals a Tully-Fisher relation with unexpectedly small scatter even for early-type galaxies. 
The morphologically-disturbed galaxies, on the other hand, showed a scatter twice as large. Galaxy interactions are likely the main origin of scatter in the T-F relation, capable of both reducing and enhancing the level of angular momentum, probably depending on the details of galaxy interaction.

We classified slow/fast rotators using the spin parameter $\lambda_{\rm R}$ and found 8 slow rotators in our sample. Approximately, $38\pm28\%$ of the slow rotators and $31\pm10\%$ of the fast rotators displayed disturbed features. The difference hints for a role of mergers on galaxy spin but is not statistically significant.

Galaxy interactions are one of the key mechanisms that change the orientation of spin of galaxies. Likewise, kinematic misalignments are more often found among the morphologically-disturbed galaxies in our sample. Kinematic misalignments are also more common among the galaxies with low angular momentum; this is probably because the same level of interaction can have a larger impact on a galaxy with lower angular momentum.

When we measure the level of kinematic perturbation, it seems necessary to account for the underlying relationship between the kinematic asymmetry and the level of rotation. We thus introduced a new parameter ($P_{\rm asym}$) which effectively reveals the impact of galaxy interactions on kinematics and photometric morphology.
 
A causality connection is mildly visible between morphological disturbance and kinematics, but not all the disturbed galaxies have unusual kinematics. Several effects may contribute to this. First, we compared the stellar kinematics within $1-2\,R_{\rm e}$ with the disturbed features in the outer regions. If we can detect stellar kinematics in the same region of galaxies where the disturbed features are shown, we might find a tighter correlation between photometry and kinematics. We also need to consider the lifetime of features in kinematics and images. Some aspects might be more quickly regularized than others through galaxy evolution (Hoffman et al. 2010; Ji et al. 2014), which could weaken correlations. 

We anticipate progress in several directions. First, it is desired to obtain and analyze the IFS and photometric data in the same extended regions of galaxies. Second, we should perform IFS observations on the numerous cluster galaxies for which deep images are already available. Although some advances have been made in this field by numerical simulations, they have so far been limited to a small number of galaxies or to semi-analytic models. But, new hydrodynamic simulations of cosmological scale will soon provide a critical clue to the evolution of spin in the hierarchical merger paradigm (e.g. Vogelsberger et al. 2014; Schaye et al. 2015; Choi et al. in prep.).


\section*{Acknowledgments}
The SAMI Galaxy Survey is based on observations made at the Anglo-Australian Telescope. The Sydney-AAO Multi-object Integral field spectrograph (SAMI) was developed jointly by the University of Sydney and the Australian Astronomical Observatory. The SAMI input catalogue is based on data taken from the Sloan Digital Sky Survey, the GAMA Survey and the VST ATLAS Survey. The SAMI Galaxy Survey is funded by the Australian Research Council Centre of Excellence for All-sky Astrophysics (CAASTRO), through project number CE110001020, and other participating institutions. The SAMI Galaxy Survey website is http://sami-survey.org/.
SO thanks DWL for his continued support.
SKY acknowledges support from the Korean National Research Foundation (NRF-2014R1A2A1A01003730).
SMC acknowledges the support of an Australian Research Council Future Fellowship (FT100100457). 
JTA acknowledges the award of a SIEF John Stocker Fellowship. 
M.S.O. acknowledges the funding support from the Australian Research Council through a Future Fellowship Fellowship (FT140100255). 
JvdS is funded under Bland-Hawthorn's ARC Laureate Fellowship (FL140100278). 
SB acknowledges the funding support from the Australian Research Council through a Future Fellowship (FT140101166).
Mahajan acknowledges the funding support from the INSPIRE Faculty award (DST/INSPIRE/04/2015/002311) and the Fast track fellowship (SB/FTP/PS-054/2013) from the Department of Science and Technology (DST) and the Science Education and Research Board (SERB) respectively.
NS acknowledges support of a University of Sydney Postdoctoral Research Fellowship.
H.J. acknowledges support from the Basic Science Research Program through the National Research Foundation of Korea (NRF), funded by the Ministry of Education (NRF-2013R1A6A3A04064993).
Support for AMM is provided by NASA through Hubble Fellowship grant \#HST-HF2-51377 awarded by the Space Telescope Science Institute, which is operated by the Association of Universities for Research in Astronomy, Inc., for NASA, under contract NAS5-26555.
This work has been supported by the Yonsei University Future-leading Research Initiative of 2015 (RMS2 2015-22-0064).

\clearpage

\appendix
\section{Deep Images}
In Figure~\ref{allimg}, we present the $r-$band deep images for the 63 galaxies used in this study. Our visual classifications are presented on each image.
    \begin{figure*}[hb]
    \centering
    \figurenum{11}
    \includegraphics[width=0.88\textwidth]{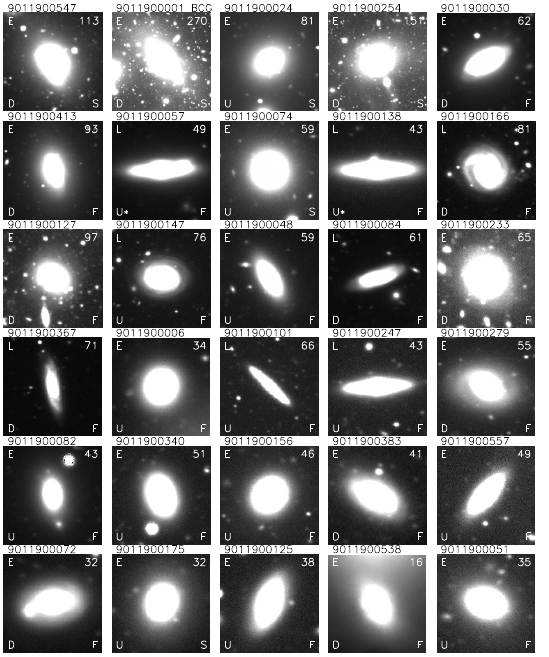}
   \caption[Sample images of morphologically-disturbed galaxies]
    {The $r$-band deep images of 63 sample galaxies. The SAMI id is presented at the top of each image. We present our visual classification on each image: The early (E) and late (L) classification is shown in the top-left corner of each image. The disturbed (D) and undisturbed (U) classification is shown in the bottom-left corner. The asterisk (*) following the U mark indicates that the galaxies have spectroscopically confirmed foreground/background contamination. The classification of fast (F) and slow (S) rotators is shown in the bottom-right corner. The number in the top-right corner shows the image size in $''$.  }
    \label{allimg}
    \end{figure*}

    \begin{figure*}[hb]
    \centering
    \figurenum{11}
    \includegraphics[width=0.88\textwidth]{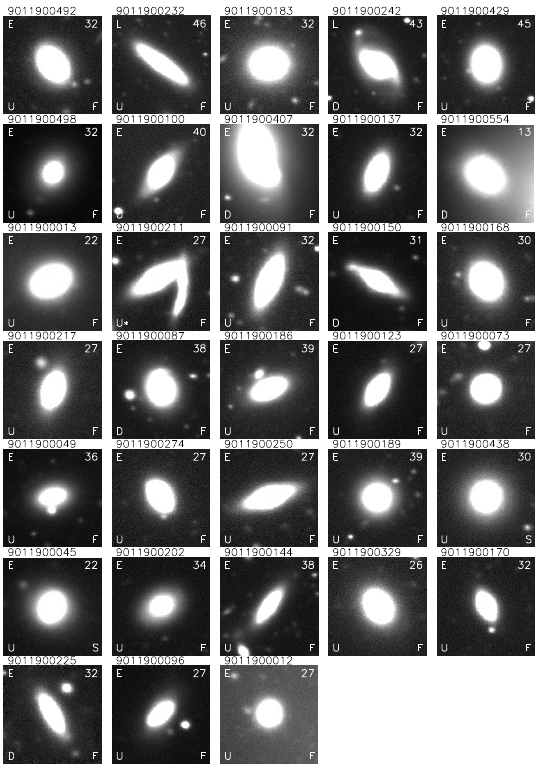}
   \caption[Sample images of morphologically-disturbed galaxies] { (continued) }
    \end{figure*}

\end{document}